\documentclass[12pt,a4paper]{article}
\usepackage{amsmath,amssymb,slashed,cancel}
\usepackage{cite,tabularx}
\usepackage{graphicx,color}
\usepackage{mathtools}
\usepackage{braket}
\usepackage{bm}

\allowdisplaybreaks

\usepackage[height=8.85in,width=6.4in]{geometry}
\renewcommand{\baselinestretch}{1.1}
\newcommand\dd{\mathrm{d}}

\definecolor{BlueViolet}{rgb}{0.2, 0.00, 0.7}
\definecolor{Blue}{rgb}{0.15, 0.00, 0.9}
\usepackage[colorlinks=true,linkcolor=Blue,citecolor=Blue,urlcolor=BlueViolet]{hyperref}

\def\thefootnote{\fnsymbol{footnote}}

\begin{document}
\begin{titlepage}
\setcounter{page}{0} 

\begin{center}

\vskip .55in

\begingroup
\centering

\hfill{\tt YITP-25-81}\\
\hfill{\tt STUPP-25-281}
\vskip .55in

{\large\bf
Gauge symmetry breaking with $S^2$ extra dimensions
}

\endgroup

\vskip .4in

{
  Kento Asai$^{\rm (a,b)}$\footnote{
  \href{mailto:kento.asai@yukawa.kyoto-u.ac.jp}
  {\tt kento.asai@yukawa.kyoto-u.ac.jp}},
  Yuki Honda$^{\rm (c)}$\footnote{
  \href{mailto:honda-yuki-wy@ynu.jp}
  {\tt honda-yuki-wy@ynu.jp}},
  Hiroki Ishikawa$^{\rm (d)}$\footnote{
  \href{mailto:h.ishikawa.404@ms.saitama-u.ac.jp}
  {\tt h.ishikawa.404@ms.saitama-u.ac.jp}},
  Joe Sato$^{\rm (c)}$\footnote{
  \href{mailto:sato-joe-mc@ynu.ac.jp}
  {\tt sato-joe-mc@ynu.ac.jp}}, \\ and
  Yasutaka Takanishi$^{\rm (d)}$\footnote{
  \href{mailto:sci56439@mail.saitama-u.ac.jp}
  {\tt sci56439@mail.saitama-u.ac.jp}}
}

\vskip 0.25in

\scalebox{0.9}{%
\begingroup\small
\begin{minipage}[t]{0.94\textwidth}
\centering\renewcommand{\arraystretch}{0.88}
\begin{tabular}{c@{\,}l}
$^{\rm(a)}$
& Yukawa Institute for Theoretical Physics, Kyoto University, 
Kyoto 606--8502, Japan \\ [1.5mm]
$^{\rm(b)}$
& Institute for Cosmic Ray Research (ICRR), The University of Tokyo, Kashiwa,\\
& Chiba 277--8582, Japan \\[1.5mm]
$^{\rm(c)}$
& Department of Physics, Facility of Engineering Science, Yokohama National University,\\
& Yokohama, Kanagawa 240--8501, Japan \\[1.5mm]
$^{\rm(d)}$
& Department of Physics, Saitama University, 255 Shimo-Okubo, Sakura-ku, \\
& Saitama 338--8570, Japan \\
\end{tabular}
\end{minipage}
\endgroup
}

\end{center}

\vskip .2in

\begin{abstract}
\noindent
We consider symmetry breaking of arbitrary gauge groups on a six-dimensional space-time which consists of a four-dimensional Minkowski space-time $M^4$ and a two-dimensional sphere $S^2$. 
We expand the gauge fields in the presence of a non-trivial background unique to $S^2$.
We analyze Kaluza-Klein(KK) modes of the gauge fields and derive the mass spectrum of the KK modes.
We found that the gauge fields (not) commuting with the background fields (do not) remain symmetry operators in four dimensions. We also discuss the mass spectrum of the extra-dimensional components of the gauge fields and identify a physical scalar $\phi$ and a Nambu-Goldstone mode $\chi$.
As a result, we obtain a method to break gauge symmetry due to the nontrivial solution for gauge fields which is a unique feature of $S^2$.

\noindent

\end{abstract}
\end{titlepage}

\begingroup
\renewcommand{\baselinestretch}{1} 
\setlength{\parskip}{2pt}          
\hrule
\tableofcontents
\vskip .2in
\hrule
\vskip .4in
\endgroup

\setcounter{page}{1}
\renewcommand{\thefootnote}{\#\arabic{footnote}}
\setcounter{footnote}{0}

\section{Introduction}
\label{sec:introduction}

The Standard Model (SM) has achieved remarkable success as a fundamental theory in particle physics and established by fitting the last piece of the puzzle, that is, discovery of the Higgs boson in 2012~\cite{ATLAS:2012yve,CMS:2012qbp}.
However, several unresolved issues remain, such as the complexity of the gauge charge of the SM fields and the theoretical origin of the Higgs particle.

The existence of multiple coupling constants arises from the fact that the SM gauge group is given by a direct product of simple groups: $G_{\mathrm{SM}} =$SU(3)$_C\times$SU(2)$_L\times$U(1)$_Y$.
Grand Unified Theory (GUT)~\cite{Georgi:1974sy,Fritzsch:1974nn,Georgi:1974my,Georgi:1975qb} provides a clear solution to this puzzle.
In GUT, a single simple gauge group $G_\mathrm{GUT}$ is assumed and broken into $G_\mathrm{SM}$ through a spontaneous breaking of the gauge symmetry, and as a result, the unification of the three gauge interactions and quantization of hypercharge are realized.

In the SM, the Higgs field plays a crucial role in the mechanism of spontaneous symmetry breaking and in giving masses to elementary particles.
Nevertheless, the origin of the Higgs field remains unknown. 
One of the promising frameworks is the Gauge-Higgs Unification (GHU) theory~\cite{Manton:1979kb,Fairlie:1979at,Fairlie:1979zy,Forgacs:1979zs,Hosotani:1983xw, Bais:1985yd,Davies:1987ei,Hosotani:1988bm,Davies:1988wt,Hatanaka:1998yp, Antoniadis:2001cv, vonGersdorff:2002as, Hasegawa:2015vqa}.  
In GHU, the SM gauge theory in the four-dimensional Minkowski spacetime is embedded in a higher-dimensional gauge theory.
The extra-dimensional components of the gauge fields behave as scalar fields in four-dimensional spacetime and can be interpreted as a Higgs field. 
Moreover, GUT has been discussed in the context of the GHU, and this kind of frameworks is called Grand Gauge-Higgs Unification~\cite{Kubo:2001zc, Burdman:2002se, Haba:2002py, Haba:2004qf, Lim:2007jv, Jittoh:2008jc, Jittoh:2008bs, Jittoh:2009th, Kojima:2011ad, Hosotani:2015hoa, Yamatsu:2015oit, Furui:2016owe, Hosotani:2017edv, Maru:2019lit, Angelescu:2021nbp, Asai:2023wxw, Maru:2024ljf, Komori:2025wji}.

In this paper, we introduce a two-dimensional sphere $S^2$ as an extra dimensional space in addition to the four-dimensional Minkowski spacetime $M^4$ and study a gauge theory on the six-dimensional spacetime. Since $S^2$ is defined as a quotient of compact Lie groups, SO(3)/SO(2), it is suitable for gauge groups and ideal for studying gauge symmetry breaking.
Indeed, the models with $S^2$ have been considered in various contexts~\cite{Horvath:1977st,Manton:1979kb,Randjbar-Daemi:1982opc,Lim:2006bx,Nomura:2008sx,Maru:2009wu,Maru:2009cu,Chiang:2010hy,Chiang:2011sj,Maru:2014cba,Dohi:2014fqa,Ueba:2019mdq,Iso:2021nas}.
Since the gauge fields on $S^2$ have a non-trivial classical solution proportional to $ \cos\theta $ as a result of the curvature, we introduce this solution as a background field~\cite{Manton:1979kb, Maru:2009wu}.
We extend the previous method and apply it to general gauge groups.
We analyze Kaluza-Klein(KK) modes expansion of the geuge fields and derive the mass spectrum of the KK modes. As a result, only the gauge fields that commute with the background field remain in four dimensions. 
Furthermore, KK modes of the extra-dimensional components of the gauge fields, which behave as scalar fields in four dimensions, are also affected by the curvature and background fields.

This paper is organized as follows. 
In section~\ref{sec:model}, we show the pure Yang-Mills theory on the six-dimensional $M^4 \times S^2$ spacetime, and introduce a nontrivial background of the gauge fields.
In section~\ref{sec:spectrum}, we analyze KK mode expansion of the gauge fields in the general framework, and derive the mass spectrum of the KK modes.
In section~\ref{sec:Application}, we show the applications for the breaking of the SU(3) and SU(5) gauge symmetries.
Finally, we summarize our work in section~\ref{sec:Summary}.

\section{\texorpdfstring{Pure Yang-Mills Theory on $M^4 \times S^2$ spacetime}{Non-Abelian Gauge Theory on M4 x S2 spacetime}}
\label{sec:model}

In this section, we review a pure Yang-Mills theory on the six-dimensional spacetime $M^4 \times S^2$ which consists of a four-dimensional Minkowski spacetime $M^4$ and a compact two-dimensional sphere $S^2$. 
We denote the indices of the coordinates of $M^4 \times S^2$ by uppercase Greek letter as $X^M (M=0,1,\cdots,5)$, and those of $M^4$ and $S^2$ by lowercase Greek letters as $X^\mu = x^\mu (\mu=0,1,2,3)$, and $X^{\hat{\alpha}} = \theta, \varphi ({\hat{\alpha}}=4,5)$, respectively.
The metric of $M^4 \times S^2$ is given by
\begin{equation}
\label{eq:metric_M6}
    G_{MN} = \begin{pmatrix}
               \eta_{\mu \nu} & 0                 \\
               0              & g_{\hat{\alpha} \hat{\beta}}
             \end{pmatrix}~,
\end{equation}
where $\eta_{\mu \nu} = {\rm diag}(1,-1,-1,-1)$ and $g_{\hat{\alpha} \hat{\beta}} = {\rm diag}(-R^2, -R^2 \sin^2\theta)$ with $R$ being the radius of $S^2$ are the metrics of $M^4$ and $S^2$, respectively.

The action of the gauge theory under consideration can be written as
\begin{align}
\label{eq:OriginalAction}
S
& = \int \, \dd^4x \, \dd\theta \, \dd\varphi \, \mathcal{L} \nonumber \\
&= \int \dd^4x \, \dd\theta \, \dd\varphi \, R^2 \sin\theta \left( -\frac{1}{2}\operatorname{Tr} F_{MN}F^{MN}\right)~,
\end{align}
where $F_{MN} = \partial_M A_N - \partial_N A_M - ig[A_M, A_N]$ with $A_M$ being the gauge field and $g$ the gauge coupling constant stands for the field strength of $A_M$.
Then, we obtain the equation of motion
\begin{equation}
\label{eq:GeneralVaccummEOMFMN}
\nabla^N F_{MN}
-ig\left[A^{N},  F_{MN}\right]= 0~,
\end{equation}
where $\nabla^M$ represents the covariant derivative of $M^4\times S^2$.
In terms of the gauge components of the Lie algebra,
\begin{equation}
\label{eq:VaccummEOMFMN}
\nabla^N F_{MNA}
+
gf_{AB}{}^C A^{NB} F_{MNC} = 0~.
\end{equation}
Here, indices $A, B, \cdots$ run over all components of the Lie algebra, and $f_{AB}{}^C$ is a structure constant of the gauge group.
By setting $M = \mu, \theta, \varphi$ in eq.~\eqref{eq:VaccummEOMFMN}, the equations of motion for the vacuum values of $A_{\mu A}, A_{\theta A}, A_{\varphi A}$ take the following form:
\begin{align}
\label{eq:VaccummEOMAmu}
   & \left( -\square + \frac{1}{R^2}{\bm{L}}^2 \right)A_{\mu A}
    + \nabla_\mu \nabla^N A_{NA} 
    + gf_{ABC} A_\mu^B \nabla^N A_N^C \nonumber \\
  & \qquad
    =-gf_{AB}{}^C A^{NB} \nabla_\mu A_{NC} + 2gf_{AB}{}^C A^{NB}\nabla_N A_{\mu C} - g^2f_{ABE}f^E{}_{CD}A^{NB} A_\mu^C A_N^D~, \\
\label{eq:VaccummEOMAtheta}
  & \left\{ -\square + \frac{1}{R^2}\left( \bm{L}^2-\frac{1}{\sin^2\theta} \right) \right\}A_{\theta A}
    - \frac{2}{R^2}\frac{\cos\theta}{\sin^3\theta}\partial_\varphi A_{\varphi A}
    + \nabla_\theta \nabla^N A_{NA}
    + gf_{ABC} A_\theta^B \nabla^N A_N^C \nonumber \\
  & \qquad
    =-gf_{AB}{}^C A^{NB} \nabla_\theta A_{NC} + 2gf_{AB}{}^C A^{NB}\nabla_N A_{\theta C} - g^2f_{ABE}f^E{}_{CD}A^{NB} A_\theta^C A_N^D~, \\
\label{eq:VaccummEOMAvarphi}
  & \frac{2}{R^2}\frac{\cos\theta}{\sin\theta} \partial_\varphi A_{\theta A}
    + \left\{ -\square + \frac{1}{R^2}\left( {\bm{L}}^2-2\frac{\cos\theta}{\sin\theta}\partial_\theta \right) \right\}A_{\varphi A}
    + \nabla_\varphi \nabla^N A_{NA}
    + gf_{ABC} A_\varphi^B \nabla^N A_N^C \nonumber \\
  & \qquad
    =-gf_{AB}{}^C A^{NB} \nabla_\varphi A_{NC} + 2gf_{AB}{}^C A^{NB}\nabla_N A_{\varphi C} - g^2f_{ABE}f^E{}_{CD}A^{NB} A_\varphi^C A_N^D~, 
\end{align}
where $\square \equiv \partial_\mu \partial^\mu$ is the d'Alembertian, and
$
\bm{L}^2\equiv-(1/\sin\theta)\partial_\theta
    \left(
    \sin\theta\partial_\theta
    \right)
    -
    ({1}/{\sin^2\theta})\partial_\varphi^2
$
is the square of angular momentum operator.
As a very remarkable feature of $S^2$, we have a non-trivial solution for eq.~\eqref{eq:GeneralVaccummEOMFMN},
\begin{equation} 
\label{eq:BackgroundCandidate} 
A_{M} = 
\begin{dcases}
    \mu\cdot H \cos\theta & (M=\varphi) \\ 
    0 & (\text{otherwise})
\end{dcases}~,
\end{equation}
where $H$ are the Cartan generators. 
Therefore, $\braket{A_{\varphi}} = \mu\cdot H \cos\theta$ can be introduced as the background field~\cite{Manton:1979kb, Maru:2009wu}.
This background fields excite the gauge field energy and hence at first glance
they must not be taken.
However, as shown in the U(1) gauge case, these kinds of background fields yield massless fermions and stabilize the entire model~\cite{Manton:1979kb, Randjbar-Daemi:1982opc, Randjbar-Daemi:1982bjy, Maru:2009wu},
and accordingly, it is reasonable to assume that a similar discussion holds in general.

\section{Mass spectrum of Kaluza--Klein modes}
\label{sec:spectrum}

In this section, we perform a KK expansion of the gauge fields in the presence of the background fields, substitute these modes into the action eq.~\eqref{eq:OriginalAction}, and derive the mass spectrum of the KK modes of the gauge field. 
Hereafter, we denote the quantum fluctuations as $A_M$ which are the deviation from the background field $\braket{A_M}$.

\subsection{Quadratic terms for gauge fields}

In this subsection, we focus on the quadratic terms of gauge field $A_M$ in the action to evaluate the mass spectrum of the gauge fields. 
We use Cartan-Weyl basis:
\begin{equation}
    A_M =
    \sum_{i:\;\textrm{all Cartans}}
        A_M^i H_i 
    + \sum_{\alpha:\;\textrm{all roots}}
        A_M^\alpha E_\alpha~,
\end{equation}
where $i, j, \cdots$ run over the Cartan subalgebra and $E_{\pm \alpha}$ are the raising and lowering operators associated with the root $\alpha$.  
In the following section, we omit the summation symbols. Whenever an index variable appears twice (or three times in some cases) in a single term, we automatically sum over all values.

Under this basis, the Lagrangian is as follows:
\begin{align}
\label{eq:QuadGauge}
  &\mathcal{L}^\textrm{quadratic}_\textrm{gauge} \nonumber \\
  &=-\frac{1}{4} R^2\sin\theta\bigg\{
   (\partial_\mu A_\nu{}_i -\partial_\nu A_\mu{}_i)(\partial^\mu A^\nu{}^i -\partial^\nu A^\mu{}^i)
   +(\partial_\mu A_\nu{}_{, -\alpha} -\partial_\nu A_\mu{}_{, -\alpha})(\partial^\mu A^\nu{}^\alpha -\partial^\nu A^\mu{}^\alpha) \nonumber \\
  &\hspace{30mm}
   -\frac{2}{R^2} (\partial_\theta A_\mu{}_i)(\partial_\theta A^\mu{}^i) -\frac{2}{R^2 \sin^2 \theta}(\partial_\varphi A_\mu{}_i (\partial_\varphi A^\mu{}^i)
   -\frac{2}{R^2} (\partial_\theta A_\mu^{-\alpha})(\partial_\theta A^{\mu\alpha}) \nonumber \\
  &\hspace{30mm}
   -\frac{2}{R^2 \sin^2 \theta}\left[(\partial_\varphi A_\mu^{-\alpha})(\partial_\varphi A^{\mu\alpha}) +2ik_\alpha\cos\theta A_\mu^{-\alpha} \partial_\varphi A^\mu{}^{\alpha}
   +k_\alpha^2 \cos^2\theta A_\mu^{-\alpha} A^\mu{}^{\alpha}\right] \nonumber \\
  &\hspace{30mm}
   -\frac{2}{R^2} \left[(\partial_\mu A_\theta{}_i)(\partial^\mu A_\theta{}^i) +(\partial_\mu A_\theta{}^{-\alpha})(\partial^\mu A_\theta{}^\alpha) \right] \nonumber \\
  &\hspace{30mm}
   -\frac{2}{R^2 \sin^2\theta}
   \left[ (\partial_\mu A_\varphi{}_i)(\partial^\mu A_\varphi{}^i)
   +(\partial_\mu A_\varphi{}^{-\alpha})(\partial^\mu A_\varphi{}^\alpha)
   \right] \nonumber \\
  &\hspace{30mm}
   +\frac{2}{R^4\sin^2\theta}(\partial_\theta A_\varphi^i -\partial_\varphi A_\theta^i)(\partial_\theta A_\varphi^i -\partial_\varphi A_\theta^i)
   +\frac{2}{R^4\sin^2\theta}(2ik_\alpha \sin\theta A_\theta^\alpha A_\varphi^{-\alpha}) \nonumber \\
  &\hspace{30mm}
   +\frac{2}{R^4\sin^2\theta}
   (\partial_\theta A_\varphi^{-\alpha} -\partial_\varphi A_\theta^{-\alpha}-ik_\alpha \cos\theta A_\theta^{-\alpha})
   (\partial_\theta A_\varphi^\alpha -\partial_\varphi A_\theta^\alpha + ik_\alpha \cos\theta A_\theta^\alpha)\bigg\} \nonumber \\
  &\hspace{1em}
   -R^2 \sin\theta \bigg\{ -\frac{1}{R^2\sin\theta} (\partial_\mu A^\mu{}_i)\left(\partial_\theta \sin\theta A_\theta^i \right) -\frac{1}{R^2\sin^2\theta}(\partial_\mu A^\mu{}_i)\left(\partial_\varphi A_\varphi^i\right) \nonumber \\
  &\hspace{26mm} 
   -\frac{1}{R^2\sin\theta}(\partial_\mu A^\mu {}^{-\alpha})\left(\partial_\theta \sin\theta A_\theta^\alpha \right)
   -\frac{1}{R^2\sin^2\theta}(\partial_\mu A^\mu{}^{-\alpha})\left(\partial_\varphi A_\varphi^\alpha\right) \nonumber \\
  &\hspace{26mm}
   +ik_\alpha\frac{\cos\theta}{R^2\sin^2\theta} (\partial_\mu A^\mu{}^{-\alpha})A_\varphi^\alpha \bigg\}~.
\end{align}
In a four-dimensional theory, the four-dimensional components $A_\mu$ behave as gauge fields, while the extra-dimensional components $A_\theta$ and $A_\varphi$ behave as scalar fields.
Therefore, in the discussion of gauge symmetry breaking, when some of the $A_\mu$ components acquire a mass, the corresponding components of $A_\theta$ and $A_\varphi$ are expected to play the role of Nambu-Goldstone bosons. 
To clarify this picture, we perform the following transformation on $A_\theta$ and $A_\varphi$:
\begin{align}
\label{eq:Transformation1}
  A_\theta^i
  &= -\frac{1}{\sin\theta}\partial_\varphi \phi^i +\partial_\theta \chi^i~,\\
\label{eq:Transformation2}
  A_\varphi^i 
  &= \sin\theta\partial_\theta \phi^i +\partial_\varphi \chi^i~, \\
\label{eq:Transformation3}
  A_\theta^\alpha 
  &= -\frac{1}{\sin\theta}\partial_\varphi \phi^\alpha +\partial_\theta \chi^\alpha +ik_\alpha \frac{\cos\theta}{\sin\theta} \phi^\alpha~ \quad\textrm{(no sum)}~, \\
\label{eq:Transformation4}
  A_\varphi^\alpha 
  &= \sin\theta\partial_\theta \phi^\alpha +\partial_\varphi \chi^\alpha - ik_\alpha \cos\theta \chi^\alpha~ \quad\textrm{(no sum)}~.
\end{align}
Here we define $k_\alpha \equiv g\alpha\cdot\mu$ as follows.
\begin{align}
    g\left[\mu\cdot H, A_M\right] 
    &= g A_M^\alpha \mu\cdot \left[H, E_\alpha\right] \nonumber\\
    &= g A_M^\alpha \mu\cdot 
    \alpha E_\alpha \nonumber\\
    &= k_\alpha A_M^\alpha E_\alpha~.
\end{align}
Therefore, $k_\alpha$ is interpreted as a charge of the $\alpha$-direction gauge field with respect to the background field.

As we will show, $\phi$ denotes the physical scalar field and $\chi$ denotes the Nambu-Goldstone boson. 
In fact, applying eqs.~\eqref{eq:Transformation1}, \eqref{eq:Transformation2}, \eqref{eq:Transformation3}, and \eqref{eq:Transformation4} to linear parts of $F_{\theta\varphi}$ in eq.~\eqref{eq:QuadGauge}, we found that it is mainly occupied by the physical scalar field $\phi$:
\begin{align}
\label{eq:LinearTermofF}
  \left.F_{\theta \varphi}\right|_\textrm{linear}
  &= ( \partial_\theta A_\varphi^i -\partial_\varphi A_\theta^i)H_i
     + ( \partial_\theta A_\varphi^\alpha -\partial_\varphi A_\theta^\alpha + i k_\alpha \cos\theta A_\theta^\alpha ) E_\alpha \nonumber \\
  &= -\sin\theta \left(\bm{L}^2 \phi^i\right) H_i 
     -\sin\theta \left( \tilde{\bm{J}}^{(\alpha)}{}^2 \phi^\alpha \right) E_\alpha
     + i k_\alpha \sin\theta \chi^\alpha E_\alpha~,
\end{align}
$\tilde{\bm{J}}^{(\alpha)}{}^2$ is an operator that has the same algebra as $\bm{L}^2$, but a different representation.
We define $\tilde{\bm{J}}^{(\alpha)}{}^2$ for each root $\alpha$
\begin{align}
\label{eq:DefinitionOfJ2}
  \tilde{\bm{J}}^{(\alpha)2}
  &\equiv 
  -\frac{1}{\sin\theta}\partial_\theta \left(   \sin\theta\partial_\theta \right)
  - \frac{1}{\sin^2\theta}\partial_\varphi^2
  + 2\frac{\cos\theta}{\sin^2\theta}k_\alpha i\partial_\varphi
  + \frac{\cos^2\theta}{\sin^2\theta} k_\alpha^2 \nonumber \\
  &=
  \bm{J}^{(\alpha)2} - k_\alpha^2~.
\end{align}
$\bm{J}^{(\alpha)2}$ are the square of the operators $J^{(\alpha)}_1$, $J^{(\alpha)}_2$, and $J^{(\alpha)}_3$, which are defined as follows:
\begin{align}
  J^{(\alpha)}_1 
  &= i \left( \sin\varphi \partial_\theta + \frac{\cos\theta}{\sin\theta} \cos\varphi \partial_\varphi \right)
  - k_\alpha \frac{\cos\varphi}{\sin\theta}~, \nonumber\\ 
  J^{(\alpha)}_2
  &= -i\left( \cos\varphi \partial_\theta - \frac{\cos\theta}{\sin\theta} \sin\varphi \partial_\varphi \right)
  - k_\alpha \frac{\sin\varphi}{\sin\theta}~, \nonumber\\
  J^{(\alpha)}_3
  &= -i\partial_\varphi~.
\end{align}
Then, $\bm{J}^{(\alpha)2}$ and $J^{(\alpha)}_i$ satisfy the SU(2) algebra:
\begin{align}
\label{eq:su2algebra}
  \bm{J}^{(\alpha)2}
  &= \sum_{i=1}^3 J^{(\alpha)}_i J^{(\alpha)}_i~, \\
  \bigl[J^{(\alpha)}_i, \,J^{(\alpha)}_j\bigr] 
  &= i \varepsilon_{ijk}J^{(\alpha)}_k~,
\end{align}
with $\varepsilon_{ijk}$ being the Levi-Civita symbol.
The eigenvalue $j$ of $\bm{J}^{(\alpha)2}$ takes non-negative integer and the eigenvalue $m$ of $J^{(\alpha)}_3$ takes $-j$, $-j+1$, $\cdots$, $j$.
As a different point from $\bm{L}^2$, $j$ is restricted to 
\begin{equation}
    j = |k_\alpha|, |k_\alpha|+1, |k_\alpha|+2, \cdots, 
    \label{eq:RelationshipJandK}
\end{equation}
which is proved in appendix~\ref{app:JacobiPolynomial}.
Therefore, $|k_\alpha|$ takes non-negative integer values.
When $k_\alpha$ is a not integer, there are no solutions. 
We interpret that the corresponding gauge field does not exist. 
Incidentally, when $k_\alpha =0$, $\bm{J}^{(\alpha)2}$ becomes $\bm{L}^2$.

We define the eigenfunctions of this representation $\mathcal{Y}_{k_\alpha jm}(\theta, \varphi)$: 
\begin{align}
    \label{eq:EquationOfJJ}
    \bm{J}^{(\alpha)2} \mathcal{Y}_{k_\alpha jm}(\theta, \varphi)
    &= j(j+1)\;\mathcal{Y}_{k_\alpha jm}(\theta, \varphi)~,
    \\
    \label{eq:EquationOfJ3}
    J^{(\alpha)}_3 \mathcal{Y}_{k_\alpha jm}(\theta, \varphi)
    &= m\;\mathcal{Y}_{k_\alpha jm}(\theta, \varphi)~.
\end{align}
The concrete form of $\mathcal{Y}_{k_\alpha jm}(\theta, \varphi)$ and the quantization of the eigenvalues are given in appendix~\ref{app:JacobiPolynomial}, where it is shown that eq.~\eqref{eq:RelationshipJandK} is ensured by the condition that the surface terms vanish. Therefore, we omit all the surface terms by partial integration hereafter.
Note that $\chi$ appears as a massless scalar with careful calculations of all terms of eq.~\eqref{eq:QuadGauge} although eq.~\eqref{eq:LinearTermofF} still have $\chi$ in root components.

We perform the following calculation to consider gauge-fixing function. 
First of all, we define the covariant derivative in the $M$-direction as
\begin{align}
  D_M A_N
  &\equiv \nabla_M A_N -ig\left[\braket{A_M}, A_N\right] \nonumber \\
  &\equiv \partial_M A_N - \Gamma_{MN}^{P} A_P -ig\left[\braket{A_M}, A_N\right]~,
\end{align}
where $\Gamma_{MN}^{P}$ are Christoffel symbols.
Using eqs.~\eqref{eq:Transformation1}, \eqref{eq:Transformation2}, \eqref{eq:Transformation3}, and \eqref{eq:Transformation4}, we obtain
\begin{align}
  D_{\hat{\alpha}} A^{\hat{\alpha}}
  &=
  \left\{
    -\frac{1}{R^2\sin\theta}\left(\partial_\theta \sin\theta A_\theta^i \right)
    -\frac{1}{R^2\sin^2\theta}\left(\partial_\varphi A_\varphi^i\right)
      \right\}H_i
  \nonumber \\
  &\qquad +
  \left\{
    -\frac{1}{R^2\sin\theta}\left(\partial_\theta \sin\theta A_\theta^\alpha \right)
    -\frac{1}{R^2\sin^2\theta}\left(\partial_\varphi A_\varphi^\alpha\right) 
    + i k_\alpha\frac{\cos\theta}{R^2\sin^2\theta} A_\varphi^\alpha 
  \right\} E_\alpha
  \nonumber \\
  &=
  \frac{1}{R^2}\left(\hat{\bm{L}}^2 \chi^i\right)H_i
  + \frac{1}{R^2}\left(\hat{\bm{J}}^{(\alpha)}{}^2 \chi^\alpha \right) E_\alpha
  + \frac{1}{R^2}ik_\alpha \phi^\alpha E_\alpha~.
\end{align}
Then, we define gauge-fixing function $G(A)$ and the gauge-fixing Lagrangian $\mathcal{L}_{\textrm{gf}}$ as follows:
\begin{align}
  G(A) 
  &=
  D_\mu A^\mu 
  +\xi\left( D_{\hat{\alpha}} A^{\hat{\alpha}} -\frac{1}{R^2} ik_\alpha \phi^\alpha E_\alpha \right)~,\\
  \mathcal{L}_{\textrm{gf}}
  &= 
  R^2 \sin\theta \left( -\frac{1}{\xi} \operatorname{tr}\left[G(A_M)^2\right] \right) \nonumber \\
  &=-\frac{1}{\xi} R^2\sin\theta\left\{  \; 
    \operatorname{tr}\left[(\partial_\mu A^\mu)^2\right]
    +2\xi\operatorname{tr}\left[(\partial_\mu A^\mu)\left(D_{\hat{\alpha}} A^{\hat{\alpha}}-\frac{1}{R^2} ik_\alpha \phi^\alpha E_\alpha\right)\right] \right. \nonumber \\
  &\left. \hspace{30mm}
    +\xi^2\operatorname{tr}\left[\left(D_{\hat{\alpha}} A^{\hat{\alpha}}-\frac{1}{R^2} ik_\alpha \phi^\alpha E_\alpha\right)^2\right] \right\}~,
    \label{eq:GaugeFixingLagrangian}
\end{align}
where $\xi$ is a gauge-fixing parameter.
Combining eqs.~\eqref{eq:QuadGauge} and \eqref{eq:GaugeFixingLagrangian} and performing partial integration, we obtain
\begin{align}
\label{eq:QuadGaugeAndGaugeFixing}
  &\mathcal{L}^\textrm{quadratic}_\textrm{gauge} + \mathcal{L}_{\textrm{gf}}
  \nonumber \\
  &= 
  -\frac{1}{4} R^2\sin\theta\bigg\{
     (\partial_\mu A_\nu{}_i -\partial_\nu A_\mu{}_i)(\partial^\mu A^\nu{}^i -\partial^\nu A^\mu{}^i)
  +(\partial_\mu A_\nu{}_{, -\alpha} -\partial_\nu A_\mu{}_{, -\alpha})(\partial^\mu A^\nu{}^\alpha -\partial^\nu A^\mu{}^\alpha)
  \nonumber \\
  &\hspace{30mm}
  -\frac{2}{R^2} (\partial_\theta A_\mu{}_i)(\partial_\theta A^\mu{}^i)
  -\frac{2}{R^2 \sin^2 \theta}(\partial_\varphi A_\mu{}_i)(\partial_\varphi A^\mu{}^i)
  -\frac{2}{R^2} (\partial_\theta A_\mu^{-\alpha})(\partial_\theta A^{\mu\alpha})
  \nonumber \\
  &\hspace{30mm}
  -\frac{2}{R^2 \sin^2 \theta}\left[(\partial_\varphi A_\mu^{-\alpha})(\partial_\varphi A^{\mu\alpha})
  +2ik_\alpha\cos\theta A_\mu^{-\alpha} \partial_\varphi A^\mu{}^{\alpha}
  +k_\alpha^2 \cos^2\theta A_\mu^{-\alpha} A^\mu{}^{\alpha}\right]
  \nonumber \\
  &\hspace{30mm}
  -\frac{2}{R^2}
  \left[(\partial_\mu A_\theta{}_i)(\partial^\mu A_\theta{}^i) 
         +(\partial_\mu A_\theta{}^{-\alpha})(\partial^\mu A_\theta{}^\alpha) 
  \right]
  \nonumber \\
  &\hspace{30mm}
  -\frac{2}{R^2 \sin^2\theta}
  \left[ (\partial_\mu A_\varphi{}_i)(\partial^\mu A_\varphi{}^i)
        +(\partial_\mu A_\varphi{}^{-\alpha})(\partial^\mu A_\varphi{}^\alpha)
  \right]
  \nonumber \\
  &\hspace{30mm}
  +\frac{2}{R^4\sin^2\theta}(\partial_\theta A_\varphi^i -\partial_\varphi A_\theta^i)(\partial_\theta A_\varphi^i -\partial_\varphi A_\theta^i)
  +\frac{2}{R^4\sin^2\theta}\left(2ik_\alpha \sin\theta A_\theta^\alpha A_\varphi^{-\alpha}\right)
  \nonumber \\
  &\hspace{30mm}
  +\frac{2}{R^4\sin^2\theta}
  (\partial_\theta A_\varphi^{-\alpha} -\partial_\varphi A_\theta^{-\alpha}-ik_\alpha \cos\theta A_\theta^{-\alpha})
  (\partial_\theta A_\varphi^\alpha -\partial_\varphi A_\theta^\alpha
   + ik_\alpha \cos\theta A_\theta^\alpha)\bigg\}
  \nonumber \\
  &\hspace{1em}
  -\frac{1}{2\xi} R^2\sin\theta
   \left\{  \; 
      (\partial_\mu A^\mu{}_i)(\partial_\nu A^\nu{}^i) 
       +(\partial_\mu A^\mu{}^{-\alpha})(\partial_\nu A^\nu{}^\alpha)
   \right\}
  \nonumber \\
  &\hspace{1em}
  - \frac{\xi}{2R^2}\sin\theta
  \left\{
     \left(\hat{\bm{L}}^2 \chi_i\right)\left(\hat{\bm{L}}^2 \chi^i\right)
     +\left( \hat{\bm{J}}^{(-\alpha)}{}^2 \chi^{-\alpha} \right)
      \left(\hat{\bm{J}}^{(\alpha)}{}^2 \chi^\alpha \right)
  \right\}
  +\sin\theta
  (\partial_\mu A^\mu{}^{-\alpha})ik_\alpha \phi^\alpha~.
\end{align}
The cross terms between $A_\mu$ and $A_{\hat{\alpha}}$ in eq.~\eqref{eq:QuadGauge} are canceled by the gauge-fixing terms,  leaving only the cross term 
$
\sin\theta\,(\partial_\mu A^{-\alpha\,\mu})\,i k_\alpha\,\phi^\alpha.
$
Since $\partial_\mu A^\mu$ will be zero in the limit of $\xi\to0$, this terms are unphysical. 
Therefore, we treat it perturbatively.

\subsection{\texorpdfstring{KK mass of $A_\mu$}{KK mass of A\_mu}}
\label{sec:gauge}

Hereafter, we consider KK expansion of the $A_\mu$ and $A_{\hat{\alpha}}$ in eq.~\eqref{eq:QuadGaugeAndGaugeFixing} and analyze mass spectrum of KK modes.

First, we consider four-dimensional gauge field $A_\mu$.
Gauge fields that do not possess massless modes correspond to the directions in which the symmetry is broken, whereas gauge fields that do possess massless modes correspond to the directions in which the symmetry remains in four dimensions.

By partial integration, we obtain
\begin{equation}
\label{eq:lagrangian2}
    \mathcal{L}_\textrm{gauge}^\textrm{quadratic}= 
    \frac{1}{2} R^2 \sin \theta
    \left[
        A_\mu^i
        \left(
        \square
        +
        \frac{1}{R^2}
        \bm{L}^2
        \right)
        A^{\mu i}
        +
        A_\mu^{-\alpha}
        \left\{
        \square
        +
        \frac{1}{R^2}
        \tilde{\bm{J}}^{(\alpha)2}
        \right\}
        A^{\mu \alpha}
        \right]~,
\end{equation}
where we omit the gauge-fixing terms since these terms proportional to $\frac{1}{\xi}$ are the same as usual gauge fixing.

Noting that the gauge fields $A_\mu$ are real fields, we can rewrite $A_\mu^\alpha$ in terms of real fields:
\begin{equation}
    A^{(\alpha)1}_\mu 
    = \frac{A_\mu^\alpha + A_\mu^{-\alpha}}{\sqrt{2}}~, 
    \qquad 
    A^{(\alpha)2}_\mu = \frac{A_\mu^\alpha - A_\mu^{-\alpha}}{-i\sqrt{2}}~.
\end{equation}
Correspondingly, we choose the following Hermitian basis
\begin{equation}
    E^{(\alpha)}_1 
    = \frac{E_\alpha + E_{-\alpha}}{\sqrt{2}}~,
    \qquad 
    E^{(\alpha)}_2 
    = \frac{E_\alpha - E_{-\alpha}}{i\sqrt{2}}~.
\end{equation}
We obtain the following relation
\begin{equation}
    A_\mu^\alpha E_\alpha + A_\mu^{-\alpha}E_{-\alpha} 
    = A^{(\alpha)1}_\mu E^{(\alpha)}_1 + A^{(\alpha)2}_\mu E^{(\alpha)}_2~.
\end{equation}
Then, eq.~\eqref{eq:lagrangian2} becomes
\begin{align}
\label{eq:QuadraticAmu}
    \mathcal{L}_\textrm{gauge}^\textrm{quadratic}
    =
    &\frac{1}{2} R^2 \sin \theta
    \left[
        A_\mu^i
        \left(
        \square
        +
        \frac{1}{R^2}
        \bm{L}^2
        \right)
        A^{\mu i} \right. 
    \nonumber \\
    &\left. \hspace{25mm}
    + \sum_{\substack{\alpha: \text{all positive roots} \\ r=1,2}}
        A^{(\alpha)r}_\mu
        \left\{
        \square
        +
        \frac{1}{R^2}
        \left(
            \tilde{\bm{J}}^{(\alpha)2}
        \right)
        \right\}
        A^{\mu (\alpha)r}
        \right]~.
\end{align}

Since the eigenfunctions of $\bm{L}^2$ are the ordinary spherical harmonics $Y_{lm}$, the KK expansion of the Cartan components $A_\mu^i$ is given by
\begin{equation}
\label{eq:KKexpansionOfAmuCartan}
    A_\mu^i (x, \theta, \varphi)
    =
    \sum_{l=0}^\infty \sum_{m=-l}^{l}\,
    \frac{\sqrt{2}}{R} \, A^{i}_{\mu, lm}(x) \,
    Y_{lm}(\theta, \varphi)~,
\end{equation}
where $l$ takes non-negative integer values $0, 1, 2, \cdots$ and $m$ takes $-l, -l+1, \cdots, l$. The corresponding KK masses are equal to
\begin{equation}
\label{eq:KKmass}
    \frac{l(l+1)}{R^2}~,
\end{equation}
and the state with $l = 0$ corresponds to the massless mode.

Since the eigenfunctions of $\tilde{\bm{J}}^{(\alpha)2}$ are $\mathcal{Y}_{k_\alpha jm}$ which we defined in the last section, the KK expansion of the root components $A_\mu^{(\alpha)r}$ is given by
\begin{equation}
\label{eq:KKexpansionOfAmuRoot}
    A_\mu^{(\alpha)r}(x, \theta, \varphi)
    =
    \sum_{j = |k_\alpha|}^\infty ~\sum_{m = -j}^j
    \frac{\sqrt{2}}{R} A^{(\alpha)r}_{\mu, jm}(x)
    \mathcal{Y}_{k_\alpha jm}(z, \varphi)~,
\end{equation} 
where $j$ takes non-negative integer values $|k_\alpha|,|k_\alpha| + 1,|k_\alpha|+ 2, \cdots$, and $m$ takes $-j, -j+1, \cdots, j$.
The KK masses are given by
\begin{equation}
    \label{eq:KKmassmod}
    \frac{j(j+1) - k_\alpha^2}{R^2}~.
\end{equation}
It is important to mention that, compared to eq.~\eqref{eq:KKmass}, the values of $j(j+1)$ are shifted to $j(j+1) - k_\alpha^2$.

By substituting eqs.~\eqref{eq:KKexpansionOfAmuCartan} and \eqref{eq:KKexpansionOfAmuRoot} into
eq.~\eqref{eq:QuadraticAmu} and
integrating over $\theta$ and $\varphi$, we obtain the kinetic and
mass terms of the four-dimensional gauge fields:
\begin{align}
  S_\mathrm{gauge}^\mathrm{quadratic}
  &=
  \int \dd^4 x
  \left[
    \sum_{l = 0}^{\infty} \sum_{m = -l}^l
    A_{\mu lm}^i
    \left\{ \square + \frac{l(l+1)}{R^2} \right\} A_{lm}^{\mu i} \right. \nonumber  \\
  &\left. \hspace{20mm} +
  \sum_{r = 1, 2}
  ~\sum_{j = |k_\alpha|}^\infty 
  ~\sum_{m = -j}^j A_{\mu jm}^{(\alpha) r}
      \left\{ \square + \frac{j(j+1) - k_\alpha^2}{R^2} \right\} A_{jm}^{\mu(\alpha) r} \right]~.
\end{align}

The gauge fields of Cartan components always have a massless mode corresponding to $l=0$. The gauge fields of root components have a massless mode when $j = k_\alpha = 0$. Therefore, when satisfying $k_\alpha = 0$, the gauge fields in the $\alpha$-direction are unbroken in four dimensions. In contrast, when $k_\alpha \neq 0$, the gauge symmetry in the $\alpha$-direction is broken.  If $k_\alpha$ is an integer, the corresponding gauge field becomes a massive vector boson; if $k_\alpha$ is non-integer, the gauge field is projected out entirely.

\subsection{\texorpdfstring{KK mass of $A_{\hat{\alpha}}$}{KK mass of A\_hat(alpha)}}
\label{sec:scalar}

Let us analyze the mass spectrum of the extra-dimensional components $A_\theta$ and $A_\varphi$. 
Using the transformation from $A_\theta$ and $A_\varphi$ to $\phi$ and $\chi$ in eqs.~\eqref{eq:Transformation1}, \eqref{eq:Transformation2}, \eqref{eq:Transformation3}, and \eqref{eq:Transformation4}, we obtain the Lagrangian as follows:
\begin{align}
    &\mathcal{L}^\textrm{quadratic}_\textrm{extra gauge} \nonumber
    \\
    &=
      -\frac{1}{2} \sin\theta
      \bigg\{
        \phi{}_i \Box \left(\bm{L}^2 \phi^i\right) + \chi{}_i \Box \left(\bm{L}^2 \chi^i\right)
        +\frac{1}{R^2}\left(\hat{\bm{L}}^2 \phi{}_i\right)\left(\hat{\bm{L}}^2 \phi^i\right)
        +\frac{\xi}{R^2}
          \left(\hat{\bm{L}}^2 \chi{}_i\right)\left(\hat{\bm{L}}^2 \chi^i\right) \nonumber
      \\
      &\hspace{21mm}
      +\phi{}^{-\alpha} \Box \left(\tilde{\bm{J}}^{(\alpha)}{}^2 \phi^{\alpha}\right) 
        +\chi{}^{-\alpha} \Box \left(\tilde{\bm{J}}^{(\alpha)}{}^2 \chi^{\alpha}\right)
        -2ik_\alpha\phi^{-\alpha}\Box \chi^\alpha
    \nonumber
    \\
      &\hspace{21mm}
        +\frac{1}{R^2}
          \left(\tilde{\bm{J}}^{(-\alpha)}{}^2 \phi^{-\alpha}\right)
          \left(\tilde{\bm{J}}^{(\alpha)}{}^2 \phi^\alpha\right)
        -\frac{1}{R^2}k_\alpha^2
        \phi^{-\alpha}\phi^{\alpha}
    \nonumber
    \\
      &\hspace{21mm}
        +\frac{\xi}{R^2}\left(\tilde{\bm{J}}^{(-\alpha)}{}^2\chi^{-\alpha}\right)
        \left(\tilde{\bm{J}}^{(\alpha)}{}^2 \chi^\alpha \right)
        \bigg\}~.
\end{align}
To realize $\phi^\alpha$ and $\chi^\alpha$ as well as $A_\mu^\alpha$, we translate them as
\begin{align}
    \phi^{(\alpha)1} = \frac{\phi^{\alpha} + \phi^{-\alpha}}{\sqrt{2}}, 
    \qquad
    \phi^{(\alpha)2} = \frac{\phi^{\alpha} - \phi^{-\alpha}}{-i\sqrt{2}},
    \label{phi realization}
    \\
    \chi^{(\alpha)1} = \frac{\chi^{\alpha} + \chi^{-\alpha}}{\sqrt{2}}, 
    \qquad
    \chi^{(\alpha)2} = \frac{\chi^{\alpha} - \chi^{-\alpha}}{-i\sqrt{2}}~.
    \label{chi realization}
\end{align}
Then, the Lagrangian are realized in the following way:
\begin{align}
\label{eq:Quadraticphichi}
    &\mathcal{L}^\textrm{quadratic}_\textrm{extra gauge} \nonumber
    \\
    &=
      -\frac{1}{2} \sin\theta
      \bigg\{
        \phi{}_i \Box \left(\bm{L}^2 \phi^i\right) + \chi{}_i \Box \left(\bm{L}^2 \chi^i\right)
        +\frac{1}{R^2}\left(\hat{\bm{L}}^2 \phi{}_i\right)\left(\hat{\bm{L}}^2 \phi^i\right)
        +\frac{\xi}{R^2}
          \left(\hat{\bm{L}}^2 \chi{}_i\right)\left(\hat{\bm{L}}^2 \chi^i\right) \nonumber
    \\
    &\hspace{22mm}+
    \sum_{\substack{\alpha: \text{all positive roots} \\ r=1,2}}\bigg[        \phi^{(\alpha)r} \Box \left(\tilde{\bm{J}}^{(\alpha)}{}^2 \phi^{(\alpha) r}\right) 
        +\chi^{(\alpha) r} \Box \left(\tilde{\bm{J}}^{(\alpha)}{}^2 \chi^{(\alpha) r}\right)
        -2ik_\alpha\phi^{(\alpha) r}\Box \chi^{(\alpha) r}
    \nonumber
    \\
      &\hspace{54mm}
        +\frac{1}{R^2}
          \left[\tilde{\bm{J}}^{(-\alpha)}{}^2 \phi^{(\alpha) r}\right]
          \left[\tilde{\bm{J}}^{(\alpha)}{}^2 \phi^{(\alpha) r}\right]
        -\frac{1}{R^2}k_\alpha^2
        \phi^{(\alpha) r}\phi^{(\alpha) r}
    \nonumber
    \\
      &\hspace{54mm}
        +\frac{\xi}{R^2}\left(\tilde{\bm{J}}^{(-\alpha)}{}^2\chi^{(\alpha) r}\right)
        \left(\tilde{\bm{J}}^{(\alpha)}{}^2 \chi^{(\alpha)}{}^r \right)
        \bigg]
        \bigg\}~.
\end{align}
The KK expansions of the Cartan components $\phi^i$ and $\chi^i$ are given by
\begin{align}
         \phi^i(x, \theta, \varphi)
        &=
        \sum_{l=0}^\infty \sum_{m=-j}^{j}\,
        \frac{\sqrt{2}}{\sqrt{l(l+1)}} \, \phi^{i}_{lm}(x) \,
        Y_{lm}(\theta, \varphi), 
    \label{eq:KKexpansionOFphiCartan}
        \\
        \chi^i(x, \theta, \varphi)
        &=
        \sum_{l=0}^\infty \sum_{m=-j}^{j}\,
        \frac{\sqrt{2}}{\sqrt{l(l+1)}} \, \chi^{i}_{lm}(x) \,
        Y_{lm}(\theta, \varphi)~. 
    \label{eq:KKexpansionOFchiCartan}
\end{align}
Also, the KK expansions of the root components $\phi^{(\alpha)r}$ and  $\chi^{(\alpha)r}$ are given by
\begin{align}
\label{eq:KKexpansionOFphiRoot}
    \phi^{(\alpha)r}(x, \theta, \varphi)
    &=
    \sum_{j = |k_\alpha|}^\infty \sum_{m = -j}^j
    \frac{\sqrt{2}}{\sqrt{j(j+1)-k_\alpha^2}} \phi^{(\alpha)r}_{jm}(x)
    \mathcal{Y}_{k_\alpha jm}(z, \varphi)~, \\
\label{eq:KKexpansionOFchiRoot}
    \chi^{(\alpha)r}(x, \theta, \varphi)
    &=
    \sum_{j = |k_\alpha|}^\infty \sum_{m = -j}^j
    \frac{\sqrt{2}}{\sqrt{j(j+1)-k_\alpha^2}} \chi^{(\alpha)r}_{jm}(x)
    \mathcal{Y}_{k_\alpha jm}(z, \varphi)~.
\end{align}
The quantum numbers $l$, $j$, $m$, and $k_\alpha$ are exactly the same as in the case of $A_\mu$.
By substituting eqs.~\eqref{eq:KKexpansionOFphiCartan}, \eqref{eq:KKexpansionOFchiCartan}, \eqref{eq:KKexpansionOFphiRoot}, and \eqref{eq:KKexpansionOFchiRoot} into
eq.~\eqref{eq:Quadraticphichi} and integrating over $\theta$ and $\varphi$, we obtain the kinetic and
mass terms of the extra-dimensional gauge fields:
\begin{align}
    &S^\text{quadratic}_\text{gauge} \nonumber \\
    & =
    \int \dd^4 x \left(
    \sum_{l = 0}^{\infty} \sum_{m = -l}^l
    \left[
    -\phi_{lm}^i
    \left\{
    \square
    +
    \frac{l(l+1)}{R^2}
    \right\}
    \phi_{lm}^i
    -\chi_{lm}^i
    \left\{
    \square
    +\xi
    \frac{l(l+1)}{R^2}
    \right\}
    \chi_{lm}^i
    \right] \right.
    \nonumber \\
    &\hspace{20mm}
    +\sum_{\substack{\alpha: \text{all positive roots} \\ r = 1, 2}}
    \sum_{j = |k_\alpha|}^\infty
    \sum_{m = -j}^j
    \left[
    -\phi_{1, jm}^{(\alpha) r}
    \left\{
    \square
    +
    \frac{\left(j(j+1) - k^2\right)^2-k^2}{R^2\left(j(j+1) - k^2\right)}
    \right\} \phi_{1, jm}^{(\alpha) r}
    \right.
    \nonumber \\
    &\hspace{71mm}
    -
    \chi_{jm}^{(\alpha) r}
    \left\{
    \square
    +\xi
    \frac{j(j+1) - k^2}{R^2}
    \right\} \chi_{jm}^{(\alpha) r} 
    \nonumber \\
    &\hspace{71mm}
    \left. \left.
    +2ik_\alpha\frac{1}{j(j+1)-k_\alpha^2}
    \phi_{jm}^{(\alpha) r}\Box \chi_{jm}^{(\alpha) r}
    \right] \right)~.
\end{align}
Then, we can understand $\phi$ is a physical scalar and $\chi$ is a Nambu-Goldstone boson. In the case $k_\alpha=0$, that is, the gauge fields commute with the background fields, the mass spectrum of $\phi^{(\alpha)r}$ coincides with that of $\phi^{i}$.
Note that since massless $\chi$ clearly plays the role of the Nambu-Goldstone boson, we treat the cross kinetic terms between $\chi^\alpha$ and $\phi^\alpha$ perturbatively. 

Since $\phi$ possesses massless scalar modes, we should consider not only the quadratic terms but also the cubic and quartic terms of the gauge fields as well as the couplings to fermion fields. Such interactions can generate radiative corrections and effective mass terms called Coleman-Weinberg potential~\cite{Coleman:1973jx}. 
Massless scalar modes of $\phi$ may therefore have important phenomenological implications for physics beyond the SM~\cite{Machacek:1983tz,Arason:1991ic,Iso:2009ss, Iso:2009nw, Farzinnia:2013pga, Carone:2013wla, Khoze:2013uia, Hashimoto:2013hta, Oda:2015gna,Chiang:2017zbz,Marzo:2018nov,Kim:2019ogz,Kierkla:2022odc,Masina:2025pnp}.

\section{Application}
\label{sec:Application}

In this section, we show the gauge symmetry breaking of SU(3) and SU(5) as illustrative examples.

\subsection{Gauge symmetry breaking of SU(3)}

\begin{figure}[tbp]
\centering
\includegraphics[width=10cm]{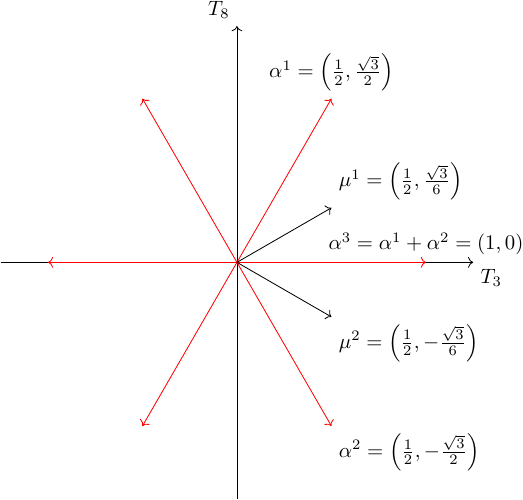}
\caption{
Diagram of the simple roots $\alpha^1$, $\alpha^2$ of SU(3) and the fundamental weights $\mu^1$, $\mu^2$. 
Here, $T_3$ and $T_8$ are the Cartan generators of SU(3).
}
\label{fig:RootDiagramOfSU3}
\end{figure}

First, we consider the case in which the gauge group is SU(3).
Let $\alpha^1, \alpha^2$ denote the simple roots of SU(3) and $\mu^1, \mu^2$ denote their fundamental weights: $\mu^i\cdot \alpha^j = \frac{1}{2}\delta^{ij}$.
Then, we define $\alpha^3 = \alpha^1+\alpha^2$ as depicted in figure~\ref{fig:RootDiagramOfSU3}.
When the background field takes the form $\mu = \mu^1-\mu^2$, we can easily calculate $k_{\alpha^1} \neq 0, k_{\alpha^2} \neq 0, k_{\alpha^3} = 0 $.
In this case, all modes of $A^{(\alpha^1)r}_{\mu,jm}$ and $A^{(\alpha^2)r}_{\mu,jm}$ are massive and the symmetries are broken, while $A^{(\alpha^3)r}_{00}$ are massless modes and the symmetry remains in the $\alpha^3$-direction. 
This implies that the higher-dimensional gauge symmetry SU(3) is spontaneously broken to SU(2)$\times$U(1) including Cartan direction.
In contrast, when $\mu = \mu^1 + \mu^2$, all $k_\alpha$ are nonzero, the symmetry is broken to only Cartan directions U(1)$\times$U(1).

\subsection{Gauge symmetry breaking of SU(5)}

Next, let the gauge group be SU(5), and denote the simple roots $ \alpha^1, \alpha^2, \alpha^3, \alpha^4$. 
We denote the fundamental weights $ \mu^1, \mu^2, \mu^3, \mu^4$ such that $\mu^i\cdot\alpha^j=\frac{1}{2} \delta^{ij}$.
When $ \mu = \mu^1 $, the vector $ \mu $ is orthogonal to $\alpha^2$, $\alpha^3$ and $ \alpha^4 $.
Therefore, the gauge group is broken to  SU(4)$\times$U(1).
On the other hand, when $ \mu = \mu^2 $, $ \mu $ is orthogonal to $ \alpha^1 $, $\alpha^3$ and $\alpha^4$.
Therefore, the gauge group is broken to SU(3)$\times$SU(2)$\times$U(1) in four dimensions.
To discuss gauge symmetry breaking of larger groups, it is convenient to expand $\mu$ with fundamental weights $\mu^i$.

\section{Summary}
\label{sec:Summary}

In this paper, we introduce a two-dimensional sphere $S^2$ as an extra dimensional space in addition to the four-dimensional Minkowski spacetime $M^4$ and analyze a gauge theory formulated on this six-dimensional spacetime.

One of the central features of this model is the nontrivial background field configuration $\braket{A_\varphi} =\mu\cdot H\cos\theta$. This setup allows us to explore mechanisms of spontaneous gauge symmetry breaking through the cuvature and topology of $S^2$, without conventional Higgs-like scalar fields.
As any function on $S^2$ can be expressed in terms of eigenfunctions of the angular momentum operators, the gauge fields commuting with the background field can be expanded in the eigenfunctions of $L_i$, while those not commuting can be expanded in the eigenfunctions of an alternative SO(3) operator, $J^{(\alpha)}_i$.
From the corresponding eigenvalues, we obtain the Kaluza-Klein mass spectrum.
The nontrivial background on $S^2$ therefore induces spontaneous gauge symmetry breaking. 
Those gauge generators which are orthogonal to the background direction remain symmetry in four dimensions.
On the contrary, those not orthogonal are broken generators. 
Additionally, performing the KK expansion of the extra-dimensional gauge components, we identify a physical scalar $\phi$ and a Nambu-Goldstone boson $\chi$.

Before conclusion, let us add a few remarks. 
The background fields raise the gauge fields energy and might therefore appear to be undesirable.
However, thanks to this background field, we could obtain massless fermions and stabilize the entire model~\cite{Manton:1979kb, Randjbar-Daemi:1982opc, Randjbar-Daemi:1982bjy, Maru:2009wu}.
Also, we obtained massless modes of $\phi$.
Then, we should consider an argument similar to Coleman-Weinberg potential~\cite{Coleman:1973jx}, in which one performs loop calculations of the scalar quartic self-interaction and the scalar-fermion coupling to consider an effective mass term. 
Based on these two considerations, we need to introduce fermions. 
We leave this study for future work.

Our framework for spontaneous gauge symmetry breaking on $S^2$ can be applied to GUT or GHU to construct explicit models that resolve the issues of the SM.

\section*{Acknowledgements}
The authors would like to thank Prof. Yoshiaki Tanii for valuable comments and suggestions.
This work was partially supported by JSPS KAKENHI Grant Numbers JP23K13097 (K.A.), JP25KJ0401 (K.A.), JP25H01524(J.S.).

\appendix
\section{\texorpdfstring{Eigenfunction of angular operator $\bm{J}^{(\alpha)}$}{Eigenfunction of angular operator J\^(alpha)}}
\label{app:JacobiPolynomial}

In this appendix, we solve eqs.~\eqref{eq:EquationOfJJ} and \eqref{eq:EquationOfJ3} under the condition that the surface terms vanish, which yields the relation eq.~\eqref{eq:RelationshipJandK} between $j$ and $k_\alpha$. Hereafter, we abbreviate $k_\alpha$ as $k$ for simplicity.

Since the eigenfunctions of $J_3^{(\alpha)} = -i\partial_\varphi$ take the form $e^{im \varphi}$ with $m = -j, -j + 1, \cdots, j$,
the eigenfunctions of $\bm{J}^{(\alpha)2}$ can be expressed as
\begin{equation}
    \mathcal{Y}_{kjm}(\theta, \varphi) = \Theta_{jkm}(\theta)e^{im\varphi} \,.
\end{equation}
By substituting this into eq.~\eqref{eq:DefinitionOfJ2} and defining $z = \cos\theta$, we obtain
\begin{equation}
    \left(
    (1-z^2)\frac{\dd^2}{\dd z^2}
    -
    2z\frac{\dd}{\dd z}
    -
    \frac{2mkz + m^2 + k^2}{1-z^2}
    +
    j(j+1)
    \right)\Theta_{jkm}
    =0~.
\end{equation}
Furthermore, by changing the variables as~\cite{Wu:1976ge, Abrikosov:2002jr}
\begin{equation}
    \Theta_{jkm} = (1-z)^\frac{|m+k|}{2} (1+z)^\frac{|m-k|}{2} w_{jkm}~,
\end{equation}
the differential equation for $w_{jkm}$ becomes
\begin{equation}
    \begin{aligned}
         & \left[
            \left(1-z^2\right) \frac{\dd^2}{\dd z^2}
            -
            \left\{
            (|m+k|+|m-k|+2) z+|m+k|-\left|m-k\right|
            \right\} \frac{\dd}{\dd z}    \right. \\
         &
            \left.
            +
            \left\{
            j - \frac{|m+k|}{2} - \frac{|m-k|}{2}
            \right\}
            \left\{
            j - \frac{|m+k|}{2} - \frac{|m-k|}{2} + |m+k| + |m-k| + 1
            \right\}
            \right] w_{jkm}=0~.
    \end{aligned}
\end{equation}

By carefully examining this equation, we observe that it corresponds to the Jacobi polynomials $P^{(\alpha, \beta)}_n(z)$ as
\begin{equation}
    \left[
        (1-z^2)\frac{\dd^2}{\dd z^2}
        -
        \{(\alpha + \beta  + 2)z + \alpha - \beta \}\frac{\dd}{\dd z}
        +
        n(n+\alpha +\beta + 1)
        \right]
    P^{(\alpha, \beta)}_n(z) = 0~,
\end{equation}
where
\begin{equation}
    \alpha = |m + k|\, ,\quad \beta = |m - k|~, \quad
    n = j - \frac{|m + k|}{2} - \frac{|m - k|}{2}~.
\end{equation}
Note that $n$ denotes the degree of the Jacobi polynomials and must be a non-negative integer to ensure the Jacobi polynomials remain finite. 
In the following, by imposing the condition that the surface term of $A_\mu$ vanishes, we demonstrate that the Jacobi polynomials must remain finite, i.e., that $n$ must be a non-negative integer. A similar argument holds for the other fields.

The Lagrangian of $A_\mu$ in eq.~\eqref{eq:lagrangian2} is followed by the surface term 
\begin{equation}
    S_{\mathrm{surface, A}_\mu}=
        \int \dd^4x \; \dd\varphi \; \left\{
      \left[\frac{1}{2}\sin\theta A_{\mu, i}(\partial_\theta A^\mu{}^{i})\right]_{\theta=0}^{\theta=\pi}
      +\left[\frac{1}{2}\sin\theta A_{\mu}^{-\alpha}(\partial_\theta A^\mu{}^{\alpha})\right]_{\theta=0}^{\theta=\pi}
       \right\}~.
\end{equation}
Expanding $A_\mu$ by eqs.~\eqref{eq:KKexpansionOfAmuCartan} and \eqref{eq:KKexpansionOfAmuRoot}, we have
\begin{align}
&-2S_{\mathrm{surface, A_\mu}}
\nonumber \\
= &
\int \dd^4x \dd\varphi \,
\left(
-
\sum_{l, l^\prime, m, m^\prime}
A_{\mu i, lm} A^{\mu i}_{l^\prime m^\prime}
N_{lm} N_{l^\prime m^\prime}
\left[
    Y_{lm} (z)
    \left( 1 - z^2 \right)
    \frac{\dd}{\dd z} Y_{l^\prime m^\prime}(z)
    \right]_{z=-1}^{z=1}
\right.
\nonumber \\
& \hspace{21mm} -
\sum_{j, j^\prime, m, m^\prime}
A^{(\alpha)}_{\mu, jm} A^{(\alpha)\mu}_{j^\prime m^\prime}
C_{kjm} C_{kj^\prime m^\prime}
\left[
(1-z)^{\frac{|m+k|}{2}}
(1+z)^{\frac{|m-k|}{2}}
P^{(|m+k|, |m-k|)}_{j-\frac{|m+k|}{2}-\frac{|m-k|}{2}} (z)
\frac{e^{im\varphi}}{\sqrt{2\pi}}
\right.
\nonumber \\
    & \hspace{21mm} \times
\left.
\left.
\left( 1-z^2 \right)
\frac{\dd}{\dd z}
\left\{
    (1-z)^{\frac{|m^\prime+k|}{2}}
    (1+z)^{\frac{|m^\prime-k|}{2}}
    P^{(|m^\prime+k|, |m^\prime-k|)}_{j^\prime-\frac{|m^\prime+k|}{2}-\frac{|m^\prime-k|}{2}} (z)
\right\}
\frac{e^{im^\prime \varphi}}{\sqrt{2\pi}}
\right]_{z=-1}^1
\right)
\nonumber \\
\equiv &
\int \dd^4x \,
\left(
- \sum_{l, l^\prime, m, m^\prime}
A_{\mu i, lm} A^{\mu i}_{l^\prime -m}
N_{lm} N_{l^\prime -m}
\sigma_{ll^\prime m}
- \sum_{j, j^\prime, m}
A^{(\alpha)}_{\mu, jm} A^{(\alpha)\mu}_{j^\prime -m}
C_{kjm} C_{kj^\prime -m}
\sigma^{(\alpha)}_{jj^\prime m}
\right)~.
\end{align}
Since $A_{\mu lm}^i$ and $A_{\mu jm}^{(\alpha)}$ are independent for each $l$ and $m$, $S_{\mathrm{surface, A}_\mu}$ can vanish if $\sigma^{(\alpha)}_{jj^\prime m} =\sigma_{ ll^\prime m} = 0$.
We use the following recurrence relation for the Jacobi polynomials:
\begin{align}
& (1-z^2) \frac{\dd P^{(a, b)}_n}{\dd z}
-
\left(
n+a+b+1
\right)
\left(
z-\frac{b-a}{2n+a+b+2}
\right)
P^{(a, b)}_n(z)
\nonumber \\
&=
-\frac{2(n+1)(n+a+b+1)}{2n+a+b+2}
P^{(a, b)}_{n+1}(z)~.
\end{align}
By setting $a = |m-k|$, $b = |m+k|$, $n = j^\prime - k^\ast$, and $k^\ast = \frac{|m+k|}{2} + \frac{|m-k|}{2}$, we can express $\sigma^{(\alpha)}_{jj^\prime m}$ by powers of $(1-z)$ and $(1+z)$, and the Jacobi function as
\begin{align}
\sigma^{(\alpha)}_{jj^\prime m}
= &
\left[
    -\frac{|m-k|}{2}
    (1-z)^{k^\ast}
    (1+z)^{k^\ast+1}
    P^{(|m+k|, |m-k|)}_{j-k^\ast} (z)
    P^{(|m-k|, |m+k|)}_{j^\prime -k^\ast} (z)
\right.
\nonumber \\
& +
\frac{|m+k|}{2}
(1-z)^{k^\ast +1}
(1+z)^{k^\ast }
P_{j-k^\ast }^{(|m+k|, |m-k|)} (z)
P_{j^\prime -k^\ast }^{(|m-k|, |m+k|)} (z)
\nonumber \\
& +
(1-z)^{k^\ast }
(1+z)^{k^\ast }
(n+a+b+1)
\left\{
    \left(z-\frac{b-a}{2 n+a+b+2}\right)
\right.
\nonumber \\
& \times 
P_{j-k^\ast }^{(|m+k|, |m-k|)} (z)
P_{j^\prime -k^\ast }^{(|m-k|, |m+k|)} (z)
-
\left.
    \left.
        \frac{2(n+1)}{2 n+a+b+2}
        P_{j-k^\ast }^{(|m+k|, |m-k|)} (z)
        P_{j^\prime -k^\ast +1}^{(|m-k|, |m+k|)} (z)
    \right\}
\right]_{z=-1}^{z=1}~.
\end{align}
Therefore, when $k^\ast \neq 0$ (i.e., $k \neq 0$ or $m \neq 0$), the Jacobi function must remain finite at $z = \pm 1$ to ensure $\sigma^{(\alpha)}_{jj^\prime m}=0$. 
In contrast, when $k^* = 0$ (i.e., $k = m = 0$), we have 
\begin{equation}
\sigma^{(\alpha)}_{j j^{\prime} 0}
=
\left(j^{\prime} - j\right)
\left\{1 + (-1)^{j + j^\prime}\right\}~,
\end{equation}
and it follows that $\sigma^{(\alpha)}_{j^\prime jm} = -\sigma^{(\alpha)}_{jj^\prime m}$.
Therefore, the surface terms automatically vanish upon summing over $j, j^\prime$.
\begin{equation}
\sum_{j, j^\prime}
\underbrace{
    \sum_{\substack{\alpha: \text{roots} \\ k_\alpha = 0}}
    A^{(\alpha)}_{\mu, j 0} A^{(\alpha)\mu}_{j^\prime 0}
    C_{kj 0} C_{kj^\prime 0}
}_{\text{symmetric under}\, j \leftrightarrow j^\prime}
\sigma^{(\alpha)}_{jj^\prime 0}
=\, 0~.
\end{equation}
Similarly, by setting $k=0$, the same argument holds for $\sigma_{ll^\prime m}$.
From the above argument, $n$ must be a non-negative integer for the surface terms to vanish. Consequently, we obtain the relation eq.~\eqref{eq:RelationshipJandK} between $j$ and $k$. Moreover, since $j$ is a non-negative integer by SU(2) algebra and the single-valuedness of the Lagrangian, $k$ must also be a non-negative integer.

The normalization constant is chosen to be
\begin{equation}
C_{jkm} =
{\sqrt{ 
\frac{2j+1}{2^{|m+k|+|m-k|+1}}
\frac{\Gamma\left(j-\frac{|m+k|}{2}-\frac{|m-k|}{2}+1\right) \Gamma\left(j+\frac{|m+k|}{2}+\frac{|m-k|}{2}
    +1\right)}{\Gamma\left(j+\frac{|m+k|}{2}-\frac{|m-k|}{2}+1\right) \Gamma\left(j-\frac{|m+k|}{2}+\frac{|m-k|}{2}+1\right)}
}}~,
\end{equation}
so that the orthogonality relation as follows, 
\begin{equation}
    \int \dd\theta \; \dd\varphi \sin\theta \;\mathcal{Y}_{k jm} \mathcal{Y}_{k j'm'}=\delta_{jj'}\delta_{mm'}~.
\end{equation}
Here, the eigenfunction $\mathcal{Y}_{kjm} (\theta, \varphi)$ is given by
\begin{equation}
\mathcal{Y}_{kjm} (\theta, \varphi)
=
C_{jkm} 
(1-z)^{\frac{|m+k|}{2}}
(1+z)^{\frac{|m-k|}{2}}
P^{(|m+k|, |m-k|)}_{j-\frac{|m+k|}{2}-\frac{|m-k|}{2}} (\cos\theta)
\frac{e^{im\varphi}}{\sqrt{2\pi}}~.
\end{equation}

\bibliographystyle{utphys28mod}
{\small
\bibliography{ref}

\providecommand{\href}[2]{#2}\begingroup\begin{thebibliography}{10}

\bibitem{ATLAS:2012yve}
{\bfseries ATLAS} Collaboration, ``{Observation of a new particle in the search
  for the Standard Model Higgs boson with the ATLAS detector at the LHC},''
  \href{https://dx.doi.org/10.1016/j.physletb.2012.08.020}{Phys.\  Lett.\  B
  {\bfseries 716} (2012) 1--29} {\ttfamily
  [\href{https://arxiv.org/abs/1207.7214}{arXiv:1207.7214}]}.

\bibitem{CMS:2012qbp}
{\bfseries CMS} Collaboration, ``{Observation of a New Boson at a Mass of 125
  GeV with the CMS Experiment at the LHC},''
  \href{https://dx.doi.org/10.1016/j.physletb.2012.08.021}{Phys.\  Lett.\  B
  {\bfseries 716} (2012) 30--61} {\ttfamily
  [\href{https://arxiv.org/abs/1207.7235}{arXiv:1207.7235}]}.

\bibitem{Georgi:1974sy}
H.~Georgi and S.~L.~Glashow, ``{Unity of All Elementary Particle Forces},''
  \href{https://dx.doi.org/10.1103/PhysRevLett.32.438}{Phys.\  Rev.\  Lett.\
  {\bfseries 32} (1974) 438--441}.

\bibitem{Fritzsch:1974nn}
H.~Fritzsch and P.~Minkowski, ``{Unified Interactions of Leptons and
  Hadrons},'' \href{https://dx.doi.org/10.1016/0003-4916(75)90211-0}{Annals
  Phys.\  {\bfseries 93} (1975) 193--266}.

\bibitem{Georgi:1974my}
H.~C. C. E.~W.~Carlson, ed., ``{The State of the Art \textemdash{} Gauge
  Theories},'' \href{https://dx.doi.org/10.1063/1.2947450}{AIP Conf.\  Proc.\
  {\bfseries 23} (1975) 575--582}.

\bibitem{Georgi:1975qb}
A.~Perlmutter and S.~M.~Widmayer, eds., ``{Unified Gauge Theories},''
  \href{https://dx.doi.org/10.1007/978-1-4613-4464-3_9}{Stud.\  Nat.\  Sci.\
  {\bfseries 9} (1975) 329--339}.

\bibitem{Manton:1979kb}
N.~S.~Manton, ``{A New Six-Dimensional Approach to the Weinberg-Salam Model},''
  \href{https://dx.doi.org/10.1016/0550-3213(79)90192-5}{Nucl.\  Phys.\  B
  {\bfseries 158} (1979) 141--153}.

\bibitem{Fairlie:1979at}
D.~B.~Fairlie, ``{Higgs' Fields and the Determination of the Weinberg Angle},''
  \href{https://dx.doi.org/10.1016/0370-2693(79)90434-9}{Phys.\  Lett.\  B
  {\bfseries 82} (1979) 97--100}.

\bibitem{Fairlie:1979zy}
D.~B.~Fairlie, ``{Two Consistent Calculations of the Weinberg Angle},''
  \href{https://dx.doi.org/10.1088/0305-4616/5/4/002}{J.\  Phys.\  G {\bfseries
  5} (1979) L55}.

\bibitem{Forgacs:1979zs}
P.~Forgacs and N.~S.~Manton, ``{Space-Time Symmetries in Gauge Theories},''
  \href{https://dx.doi.org/10.1007/BF01200108}{Commun.\  Math.\  Phys.\
  {\bfseries 72} (1980) 15}.

\bibitem{Hosotani:1983xw}
Y.~Hosotani, ``{Dynamical Mass Generation by Compact Extra Dimensions},''
  \href{https://dx.doi.org/10.1016/0370-2693(83)90170-3}{Phys.\  Lett.\  B
  {\bfseries 126} (1983) 309--313}.

\bibitem{Bais:1985yd}
F.~A.~Bais, K.~J.~Barnes, P.~Forgacs, and G.~Zoupanos, ``{Dimensional Reduction
  of Gauge Theories Yielding Unified Models Spontaneously Broken to $SU(3)
  \times U(1)$},''
  \href{https://dx.doi.org/10.1016/0550-3213(86)90274-9}{Nucl.\  Phys.\  B
  {\bfseries 263} (1986) 557--590}.

\bibitem{Davies:1987ei}
A.~T.~Davies and A.~McLachlan, ``{Gauge Group Breaking By Wilson Loops},''
  \href{https://dx.doi.org/10.1016/0370-2693(88)90776-9}{Phys.\  Lett.\  B
  {\bfseries 200} (1988) 305--311}.

\bibitem{Hosotani:1988bm}
Y.~Hosotani, ``{Dynamics of Nonintegrable Phases and Gauge Symmetry
  Breaking},'' \href{https://dx.doi.org/10.1016/0003-4916(89)90015-8}{Annals
  Phys.\  {\bfseries 190} (1989) 233}.

\bibitem{Davies:1988wt}
A.~T.~Davies and A.~McLachlan, ``{Congruency Class Effects in the Hosotani
  Model},'' \href{https://dx.doi.org/10.1016/0550-3213(89)90569-5}{Nucl.\
  Phys.\  B {\bfseries 317} (1989) 237}.

\bibitem{Hatanaka:1998yp}
H.~Hatanaka, T.~Inami, and C.~S.~Lim, ``{The Gauge hierarchy problem and higher
  dimensional gauge theories},''
  \href{https://dx.doi.org/10.1142/S021773239800276X}{Mod.\  Phys.\  Lett.\  A
  {\bfseries 13} (1998) 2601--2612} {\ttfamily
  [\href{https://arxiv.org/abs/hep-th/9805067}{hep-th/9805067}]}.

\bibitem{Antoniadis:2001cv}
I.~Antoniadis, K.~Benakli, and M.~Quiros, ``{Finite Higgs mass without
  supersymmetry},'' \href{https://dx.doi.org/10.1088/1367-2630/3/1/320}{New J.\
   Phys.\  {\bfseries 3} (2001) 20} {\ttfamily
  [\href{https://arxiv.org/abs/hep-th/0108005}{hep-th/0108005}]}.

\bibitem{vonGersdorff:2002as}
G.~von Gersdorff, N.~Irges, and M.~Quiros, ``{Bulk and brane radiative effects
  in gauge theories on orbifolds},''
  \href{https://dx.doi.org/10.1016/S0550-3213(02)00395-4}{Nucl.\  Phys.\  B
  {\bfseries 635} (2002) 127--157} {\ttfamily
  [\href{https://arxiv.org/abs/hep-th/0204223}{hep-th/0204223}]}.

\bibitem{Hasegawa:2015vqa}
K.~Hasegawa, C.~S.~Lim, and N.~Maru, ``{Predictions of the Higgs mass and the
  weak mixing angle in the 6D gauge-Higgs unification},''
  \href{https://dx.doi.org/10.7566/JPSJ.85.074101}{J.\  Phys.\  Soc.\  Jap.\
  {\bfseries 85} (2016) 074101} {\ttfamily
  [\href{https://arxiv.org/abs/1509.04818}{arXiv:1509.04818}]}.

\bibitem{Kubo:2001zc}
M.~Kubo, C.~S.~Lim, and H.~Yamashita, ``{The Hosotani mechanism in bulk gauge
  theories with an orbifold extra space $S^1/Z_2$},''
  \href{https://dx.doi.org/10.1142/S0217732302008988}{Mod.\  Phys.\  Lett.\  A
  {\bfseries 17} (2002) 2249--2264} {\ttfamily
  [\href{https://arxiv.org/abs/hep-ph/0111327}{hep-ph/0111327}]}.

\bibitem{Burdman:2002se}
G.~Burdman and Y.~Nomura, ``{Unification of Higgs and Gauge Fields in Five
  Dimensions},'' \href{https://dx.doi.org/10.1016/S0550-3213(03)00088-9}{Nucl.\
   Phys.\  B {\bfseries 656} (2003) 3--22} {\ttfamily
  [\href{https://arxiv.org/abs/hep-ph/0210257}{hep-ph/0210257}]}.

\bibitem{Haba:2002py}
N.~Haba, M.~Harada, Y.~Hosotani, and Y.~Kawamura, ``{Dynamical rearrangement of
  gauge symmetry on the orbifold S1 / Z(2)},''
  \href{https://dx.doi.org/10.1016/S0550-3213(03)00142-1}{Nucl.\  Phys.\  B
  {\bfseries 657} (2003) 169--213} {\ttfamily
  [\href{https://arxiv.org/abs/hep-ph/0212035}{hep-ph/0212035}]}. [Erratum:
  Nucl.Phys.B 669, 381--382 (2003)].

\bibitem{Haba:2004qf}
N.~Haba, Y.~Hosotani, Y.~Kawamura, and T.~Yamashita, ``{Dynamical symmetry
  breaking in gauge Higgs unification on orbifold},''
  \href{https://dx.doi.org/10.1103/PhysRevD.70.015010}{Phys.\  Rev.\  D
  {\bfseries 70} (2004) 015010} {\ttfamily
  [\href{https://arxiv.org/abs/hep-ph/0401183}{hep-ph/0401183}]}.

\bibitem{Lim:2007jv}
C.~S.~Lim and N.~Maru, ``{Towards a realistic grand gauge-Higgs unification},''
  \href{https://dx.doi.org/10.1016/j.physletb.2007.07.053}{Phys.\  Lett.\  B
  {\bfseries 653} (2007) 320--324} {\ttfamily
  [\href{https://arxiv.org/abs/0706.1397}{arXiv:0706.1397}]}.

\bibitem{Jittoh:2008jc}
T.~Jittoh, M.~Koike, T.~Nomura, J.~Sato, and T.~Shimomura, ``{Model building by
  Coset space dimensional reduction scheme using ten-dimensional coset
  spaces},'' \href{https://dx.doi.org/10.1143/PTP.120.1041}{Prog.\  Theor.\
  Phys.\  {\bfseries 120} (2008) 1041--1063} {\ttfamily
  [\href{https://arxiv.org/abs/0803.0641}{arXiv:0803.0641}]}.

\bibitem{Jittoh:2008bs}
T.~Jittoh, M.~Koike, T.~Nomura, J.~Sato, and T.~Shimomura, ``{Model building by
  coset space dimensional reduction in ten-dimensions with direct product gauge
  symmetry},'' \href{https://dx.doi.org/10.1103/PhysRevD.79.056004}{Phys.\
  Rev.\  D {\bfseries 79} (2009) 056004} {\ttfamily
  [\href{https://arxiv.org/abs/0812.0910}{arXiv:0812.0910}]}.

\bibitem{Jittoh:2009th}
T.~Jittoh, M.~Koike, T.~Nomura, J.~Sato, and Y.~Toyama, ``{Model building by
  coset space dimensional reduction in eight-dimensions},''
  \href{https://dx.doi.org/10.1016/j.physletb.2009.04.044}{Phys.\  Lett.\  B
  {\bfseries 675} (2009) 450--454} {\ttfamily
  [\href{https://arxiv.org/abs/0903.2164}{arXiv:0903.2164}]}.

\bibitem{Kojima:2011ad}
K.~Kojima, K.~Takenaga, and T.~Yamashita, ``{Grand Gauge-Higgs Unification},''
  \href{https://dx.doi.org/10.1103/PhysRevD.84.051701}{Phys.\  Rev.\  D
  {\bfseries 84} (2011) 051701} {\ttfamily
  [\href{https://arxiv.org/abs/1103.1234}{arXiv:1103.1234}]}.

\bibitem{Hosotani:2015hoa}
Y.~Hosotani and N.~Yamatsu, ``{Gauge\textendash{}Higgs grand unification},''
  \href{https://dx.doi.org/10.1093/ptep/ptv153}{PTEP {\bfseries 2015} (2015)
  111B01} {\ttfamily
  [\href{https://arxiv.org/abs/1504.03817}{arXiv:1504.03817}]}.

\bibitem{Yamatsu:2015oit}
N.~Yamatsu, ``{Gauge coupling unification in gauge\textendash{}Higgs grand
  unification},'' \href{https://dx.doi.org/10.1093/ptep/ptw023}{PTEP {\bfseries
  2016} (2016) 043B02} {\ttfamily
  [\href{https://arxiv.org/abs/1512.05559}{arXiv:1512.05559}]}.

\bibitem{Furui:2016owe}
A.~Furui, Y.~Hosotani, and N.~Yamatsu, ``{Toward Realistic Gauge-Higgs Grand
  Unification},'' \href{https://dx.doi.org/10.1093/ptep/ptw116}{PTEP {\bfseries
  2016} (2016) 093B01} {\ttfamily
  [\href{https://arxiv.org/abs/1606.07222}{arXiv:1606.07222}]}.

\bibitem{Hosotani:2017edv}
Y.~Hosotani and N.~Yamatsu, ``{Electroweak symmetry breaking and mass spectra
  in six-dimensional gauge\textendash{}Higgs grand unification},''
  \href{https://dx.doi.org/10.1093/ptep/ptx175}{PTEP {\bfseries 2018} (2018)
  023B05} {\ttfamily
  [\href{https://arxiv.org/abs/1710.04811}{arXiv:1710.04811}]}.

\bibitem{Maru:2019lit}
N.~Maru and Y.~Yatagai, ``{Fermion Mass Hierarchy in Grand Gauge-Higgs
  Unification},'' \href{https://dx.doi.org/10.1093/ptep/ptz083}{PTEP {\bfseries
  2019} (2019) 083B03} {\ttfamily
  [\href{https://arxiv.org/abs/1903.08359}{arXiv:1903.08359}]}.

\bibitem{Angelescu:2021nbp}
A.~Angelescu, A.~Bally, S.~Blasi, and F.~Goertz, ``{Minimal SU(6) gauge-Higgs
  grand unification},''
  \href{https://dx.doi.org/10.1103/PhysRevD.105.035026}{Phys.\  Rev.\  D
  {\bfseries 105} (2022) 035026} {\ttfamily
  [\href{https://arxiv.org/abs/2104.07366}{arXiv:2104.07366}]}.

\bibitem{Asai:2023wxw}
K.~Asai, J.~Sato, R.~Suda, Y.~Takanishi, and M.~J.~S.~Yang, ``{Model building
  by coset space dimensional reduction scheme using eight-dimensional coset
  spaces},'' \href{https://dx.doi.org/10.1007/JHEP11(2023)213}{JHEP {\bfseries
  11} (2023) 213} {\ttfamily
  [\href{https://arxiv.org/abs/2305.01421}{arXiv:2305.01421}]}.

\bibitem{Maru:2024ljf}
N.~Maru and R.~Nago, ``{New models of SU(6) grand gauge-Higgs unification},''
  \href{https://dx.doi.org/10.1007/JHEP11(2024)035}{JHEP {\bfseries 11} (2024)
  035} {\ttfamily [\href{https://arxiv.org/abs/2405.07463}{arXiv:2405.07463}]}.

\bibitem{Komori:2025wji}
Y.~Komori and N.~Maru, ``{SU(7) Grand Gauge-Higgs Unification}.'' {\ttfamily
  \href{https://arxiv.org/abs/2503.04090}{arXiv:2503.04090}}.

\bibitem{Horvath:1977st}
Z.~Horvath, L.~Palla, E.~Cremmer, and J.~Scherk, ``{Grand Unified Schemes and
  Spontaneous Compactification},''
  \href{https://dx.doi.org/10.1016/0550-3213(77)90351-0}{Nucl.\  Phys.\  B
  {\bfseries 127} (1977) 57--65}.

\bibitem{Randjbar-Daemi:1982opc}
S.~Randjbar-Daemi, A.~Salam, and J.~A.~Strathdee, ``{Spontaneous
  Compactification in Six-Dimensional Einstein-Maxwell Theory},''
  \href{https://dx.doi.org/10.1016/0550-3213(83)90247-X}{Nucl.\  Phys.\  B
  {\bfseries 214} (1983) 491--512}.

\bibitem{Lim:2006bx}
C.~S.~Lim, N.~Maru, and K.~Hasegawa, ``{Six Dimensional Gauge-Higgs Unification
  with an Extra Space $S^2$ and the Hierarchy Problem},''
  \href{https://dx.doi.org/10.1143/JPSJ.77.074101}{J.\  Phys.\  Soc.\  Jap.\
  {\bfseries 77} (2008) 074101} {\ttfamily
  [\href{https://arxiv.org/abs/hep-th/0605180}{hep-th/0605180}]}.

\bibitem{Nomura:2008sx}
T.~Nomura and J.~Sato, ``{Standard(-like) Model from an SO(12) Grand Unified
  Theory in six-dimensions with $S_2$ extra-space},''
  \href{https://dx.doi.org/10.1016/j.nuclphysb.2008.11.017}{Nucl.\  Phys.\  B
  {\bfseries 811} (2009) 109--122} {\ttfamily
  [\href{https://arxiv.org/abs/0810.0898}{arXiv:0810.0898}]}.

\bibitem{Maru:2009wu}
N.~Maru, T.~Nomura, J.~Sato, and M.~Yamanaka, ``{The Universal Extra
  Dimensional Model with $S^2/Z_2$ extra-space},''
  \href{https://dx.doi.org/10.1016/j.nuclphysb.2009.11.023}{Nucl.\  Phys.\  B
  {\bfseries 830} (2010) 414--433} {\ttfamily
  [\href{https://arxiv.org/abs/0904.1909}{arXiv:0904.1909}]}.

\bibitem{Maru:2009cu}
N.~Maru, T.~Nomura, J.~Sato, and M.~Yamanaka, ``{Higgs Production via Gluon
  Fusion in a Six Dimensional Universal Extra Dimension Model on $S^2/Z_2$},''
  \href{https://dx.doi.org/10.1140/epjc/s10052-010-1236-3}{Eur.\  Phys.\  J.\
  C {\bfseries 66} (2010) 283--287} {\ttfamily
  [\href{https://arxiv.org/abs/0905.4554}{arXiv:0905.4554}]}.

\bibitem{Chiang:2010hy}
C.-W.~Chiang and T.~Nomura, ``{A Six-dimensional gauge-Higgs unification model
  based on $E_6$ gauge symmetry},''
  \href{https://dx.doi.org/10.1016/j.nuclphysb.2010.09.003}{Nucl.\  Phys.\  B
  {\bfseries 842} (2011) 362--382} {\ttfamily
  [\href{https://arxiv.org/abs/1006.4446}{arXiv:1006.4446}]}.

\bibitem{Chiang:2011sj}
C.-W.~Chiang, T.~Nomura, and J.~Sato, ``{Gauge-Higgs unification models in six
  dimensions with $S^2/Z_2$ extra space and GUT gauge symmetry},''
  \href{https://dx.doi.org/10.1155/2012/260848}{Adv.\  High Energy Phys.\
  {\bfseries 2012} (2012) 260848} {\ttfamily
  [\href{https://arxiv.org/abs/1109.5835}{arXiv:1109.5835}]}.

\bibitem{Maru:2014cba}
N.~Maru, T.~Nomura, and J.~Sato, ``{One-loop radiative correction to
  Kaluza-Klein masses in $S^2/Z_2$ universal extra-dimensional model},''
  \href{https://dx.doi.org/10.1093/ptep/ptu113}{PTEP {\bfseries 2014} (2014)
  083B04} {\ttfamily
  [\href{https://arxiv.org/abs/1401.7204}{arXiv:1401.7204}]}.

\bibitem{Dohi:2014fqa}
H.~Dohi, T.~Kakuda, K.~Nishiwaki, K.-y.~Oda, and N.~Okuda, ``{Notes on
  sphere-based universal extra dimensions},'' Afr.\  Rev.\  Phys.\  {\bfseries
  9} (2014) 0069 {\ttfamily
  [\href{https://arxiv.org/abs/1406.1954}{arXiv:1406.1954}]}.

\bibitem{Ueba:2019mdq}
I.~Ueba, ``{Extended supersymmetry with central charges in Dirac action with
  curved extra dimensions},''
  \href{https://dx.doi.org/10.1103/PhysRevD.100.105001}{Phys.\  Rev.\  D
  {\bfseries 100} (2019) 105001} {\ttfamily
  [\href{https://arxiv.org/abs/1905.11673}{arXiv:1905.11673}]}.

\bibitem{Iso:2021nas}
S.~Iso, N.~Kitazawa, and T.~Suyama, ``{Gauge symmetry restoration by Higgs
  condensation in flux compactifications on coset spaces},''
  \href{https://dx.doi.org/10.1103/PhysRevD.105.045008}{Phys.\  Rev.\  D
  {\bfseries 105} (2022) 045008} {\ttfamily
  [\href{https://arxiv.org/abs/2111.06508}{arXiv:2111.06508}]}.

\bibitem{Randjbar-Daemi:1982bjy}
S.~Randjbar-Daemi and R.~Percacci, ``{Spontaneous Compactification of a
  (4+$d$)-dimensional {Kaluza-Klein} Theory Into $M_4 \times G/H$ for Arbitrary
  $G$ and $H$},'' \href{https://dx.doi.org/10.1016/0370-2693(82)90869-3}{Phys.\
   Lett.\  B {\bfseries 117} (1982) 41}.

\bibitem{Coleman:1973jx}
S.~R.~Coleman and E.~J.~Weinberg, ``{Radiative Corrections as the Origin of
  Spontaneous Symmetry Breaking},''
  \href{https://dx.doi.org/10.1103/PhysRevD.7.1888}{Phys.\  Rev.\  D {\bfseries
  7} (1973) 1888--1910}.

\bibitem{Machacek:1983tz}
M.~E.~Machacek and M.~T.~Vaughn, ``{Two Loop Renormalization Group Equations in
  a General Quantum Field Theory. 1. Wave Function Renormalization},''
  \href{https://dx.doi.org/10.1016/0550-3213(83)90610-7}{Nucl.\  Phys.\  B
  {\bfseries 222} (1983) 83--103}.

\bibitem{Arason:1991ic}
H.~Arason, D.~J.~Castano, B.~Keszthelyi, S.~Mikaelian, \emph{et al}.,
  ``{Renormalization group study of the standard model and its extensions. 1.
  The Standard model},''
  \href{https://dx.doi.org/10.1103/PhysRevD.46.3945}{Phys.\  Rev.\  D
  {\bfseries 46} (1992) 3945--3965}.

\bibitem{Iso:2009ss}
S.~Iso, N.~Okada, and Y.~Orikasa, ``{Classically conformal $B-L$ extended
  Standard Model},''
  \href{https://dx.doi.org/10.1016/j.physletb.2009.04.046}{Phys.\  Lett.\  B
  {\bfseries 676} (2009) 81--87} {\ttfamily
  [\href{https://arxiv.org/abs/0902.4050}{arXiv:0902.4050}]}.

\bibitem{Iso:2009nw}
S.~Iso, N.~Okada, and Y.~Orikasa, ``{The minimal $B-L$ model naturally realized
  at TeV scale},'' \href{https://dx.doi.org/10.1103/PhysRevD.80.115007}{Phys.\
  Rev.\  D {\bfseries 80} (2009) 115007} {\ttfamily
  [\href{https://arxiv.org/abs/0909.0128}{arXiv:0909.0128}]}.

\bibitem{Farzinnia:2013pga}
A.~Farzinnia, H.-J.~He, and J.~Ren, ``{Natural Electroweak Symmetry Breaking
  from Scale Invariant Higgs Mechanism},''
  \href{https://dx.doi.org/10.1016/j.physletb.2013.09.060}{Phys.\  Lett.\  B
  {\bfseries 727} (2013) 141--150} {\ttfamily
  [\href{https://arxiv.org/abs/1308.0295}{arXiv:1308.0295}]}.

\bibitem{Carone:2013wla}
C.~D.~Carone and R.~Ramos, ``{Classical scale-invariance, the electroweak scale
  and vector dark matter},''
  \href{https://dx.doi.org/10.1103/PhysRevD.88.055020}{Phys.\  Rev.\  D
  {\bfseries 88} (2013) 055020} {\ttfamily
  [\href{https://arxiv.org/abs/1307.8428}{arXiv:1307.8428}]}.

\bibitem{Khoze:2013uia}
V.~V.~Khoze, ``{Inflation and Dark Matter in the Higgs Portal of Classically
  Scale Invariant Standard Model},''
  \href{https://dx.doi.org/10.1007/JHEP11(2013)215}{JHEP {\bfseries 11} (2013)
  215} {\ttfamily [\href{https://arxiv.org/abs/1308.6338}{arXiv:1308.6338}]}.

\bibitem{Hashimoto:2013hta}
M.~Hashimoto, S.~Iso, and Y.~Orikasa, ``{Radiative symmetry breaking at the
  Fermi scale and flat potential at the Planck scale},''
  \href{https://dx.doi.org/10.1103/PhysRevD.89.016019}{Phys.\  Rev.\  D
  {\bfseries 89} (2014) 016019} {\ttfamily
  [\href{https://arxiv.org/abs/1310.4304}{arXiv:1310.4304}]}.

\bibitem{Oda:2015gna}
S.~Oda, N.~Okada, and D.-s.~Takahashi, ``{Classically conformal U(1)' extended
  standard model and Higgs vacuum stability},''
  \href{https://dx.doi.org/10.1103/PhysRevD.92.015026}{Phys.\  Rev.\  D
  {\bfseries 92} (2015) 015026} {\ttfamily
  [\href{https://arxiv.org/abs/1504.06291}{arXiv:1504.06291}]}.

\bibitem{Chiang:2017zbz}
C.-W.~Chiang and E.~Senaha, ``{On gauge dependence of gravitational waves from
  a first-order phase transition in classical scale-invariant $U(1)'$
  models},'' \href{https://dx.doi.org/10.1016/j.physletb.2017.09.064}{Phys.\
  Lett.\  B {\bfseries 774} (2017) 489--493} {\ttfamily
  [\href{https://arxiv.org/abs/1707.06765}{arXiv:1707.06765}]}.

\bibitem{Marzo:2018nov}
C.~Marzo, L.~Marzola, and V.~Vaskonen, ``{Phase transition and vacuum stability
  in the classically conformal B\textendash{}L model},''
  \href{https://dx.doi.org/10.1140/epjc/s10052-019-7076-x}{Eur.\  Phys.\  J.\
  C {\bfseries 79} (2019) 601} {\ttfamily
  [\href{https://arxiv.org/abs/1811.11169}{arXiv:1811.11169}]}.

\bibitem{Kim:2019ogz}
Y.~G.~Kim, K.~Y.~Lee, and S.-H.~Nam, ``{Conformal invariance and singlet
  fermionic dark matter},''
  \href{https://dx.doi.org/10.1103/PhysRevD.100.075038}{Phys.\  Rev.\  D
  {\bfseries 100} (2019) 075038} {\ttfamily
  [\href{https://arxiv.org/abs/1906.03390}{arXiv:1906.03390}]}.

\bibitem{Kierkla:2022odc}
M.~Kierkla, A.~Karam, and B.~Swiezewska, ``{Conformal model for gravitational
  waves and dark matter: a status update},''
  \href{https://dx.doi.org/10.1007/JHEP03(2023)007}{JHEP {\bfseries 03} (2023)
  007} {\ttfamily [\href{https://arxiv.org/abs/2210.07075}{arXiv:2210.07075}]}.

\bibitem{Masina:2025pnp}
I.~Masina and M.~Quiros, ``{An introduction to effective potential methods in
  field theory}.'' {\ttfamily
  \href{https://arxiv.org/abs/2501.12713}{arXiv:2501.12713}}.

\bibitem{Wu:1976ge}
T.~T.~Wu and C.~N.~Yang, ``{Dirac Monopole Without Strings: Monopole
  Harmonics},'' \href{https://dx.doi.org/10.1016/0550-3213(76)90143-7}{Nucl.\
  Phys.\  B {\bfseries 107} (1976) 365}.

\bibitem{Abrikosov:2002jr}
A.~A.~Abrikosov, Jr., ``{Dirac operator on the Riemann sphere}.'' {\ttfamily
  \href{https://arxiv.org/abs/hep-th/0212134}{hep-th/0212134}}.

\end{thebibliography}\endgroup
}

\end{document}